\documentclass[]{spie}  

\usepackage{amsmath,amsfonts,amssymb}
\usepackage{graphicx}
\usepackage[colorlinks=true, allcolors=blue]{hyperref}
\usepackage{multirow}
\usepackage{soul}
\usepackage[]{xcolor}

\definecolor{nlGreen}{rgb}{0.0, 0.7, 0.0}

\definecolor{cgHotPink}{rgb}{1.0, 0.0, 0.5}
\definecolor{jioBlue}{rgb}{0.1, 0.1, 1.0}
\definecolor{cgBlack}{rgb}{0.0, 0.0, 0.0}

\title{A deep learning model for brain vessel segmentation\\ in 3DRA with arteriovenous malformations}

\author[a,b]{Camila Garc\'ia}
\author[c]{Yibin Fang}
\author[d]{Jianmin Liu}
\author[e]{Ana Paula Narata}
\author[a,b]{Jos\'e Ignacio Orlando}
\author[a,b]{Ignacio Larrabide}

\affil[a]{Yatiris Group, PLADEMA Institute, UNICEN, Campus Universitario, Tandil, Argentina}
\affil[b]{Consejo Nacional de Investigaciones Cient\'ificas y T\'ecnicas, CONICET, Tandil, Argentina}
\affil[c]{Department of Neurovascular Disease, Shanghai Fourth People's Hospital, School of Medicine, Tongji University, Shanghai, China}
\affil[d]{Department of Neurosurgery, Changhai Hospital, Naval Medical University, Shanghai, China}
\affil[e]{Department of Neuroradiology, University Hospital of Southampton, Southampton, UK}

\authorinfo{Corresponding author contact information: \\C. Garcia: E-mail: cgarcialam@pladema.exa.unicen.edu.ar}

\begin{document} 
\maketitle

\begin{abstract} 
Segmentation of brain arterio-venous malformations (bAVMs) in 3D rotational angiographies (3DRA) is still an open problem in the literature, with high relevance for clinical practice. While deep learning models have been applied for segmenting the brain vasculature in these images, they have never been used in cases with bAVMs. This is likely caused by the difficulty to obtain sufficiently annotated data to train these approaches. In this paper we introduce a first deep learning model for blood vessel segmentation in 3DRA images of patients with bAVMs. To this end, we densely annotated 5 3DRA volumes of bAVM cases and used these to train two alternative 3DUNet-based architectures with different segmentation objectives. Our results show that the networks reach a comprehensive coverage of relevant structures for bAVM analysis, much better than what is obtained using standard methods. This is promising for achieving a better topological and morphological characterisation of the bAVM structures of interest. Furthermore, the models have the ability to segment venous structures even when missing in the ground truth labelling, which is relevant for planning interventional treatments. Ultimately, these results could be used as more reliable first initial guesses, alleviating the cumbersome task of creating manual labels. 
\end{abstract}


\section{Introduction} 
\label{sec:intro}  



Brain arteriovenous malformations (bAVMs) are an entanglement of abnormal vessels, produced when dilated feeding arteries shunt blood directly into arterialised draining veins\cite{dumont2017brain, rutledge2017brain}. If untreated, these pathological alterations have an associated risk of hemorrhages, with significant morbidity and mortality rates\cite{rutledge2014hemorrhage}. Indications for bAVMs treatment vary according to its classification---depending on factors such as the nidus size, number of feeding arteries and draining veins (hereinafter called structures of interest) and its anatomical location---and on whether it is ruptured or not \cite{spetzler1986proposed, lawton2010supplementary}. 

Different imaging protocols are used depending on the clinical scenario \cite{tranvinh2017contemporary}. In non-traumatic settings, computed tomography angiography (CTA) or magnetic resonance angiography (MRA) are frequently used as initial 3D examinations, since they do not require catheterisation. For intraoperative assistance, 3D rotational angiography (3DRA) provides high-resolution images of the vascular anatomy of the bAVM. In this study we focus on this particular modality, which allows for a clearer definition of the structures of interest \cite{dumont2017brain}.

Determining the morphology and topology of a bAVM using 3DRA allows to efficiently extract vital information for clinicians, such as size, shape, location and number of each of the structures of interest. 
The adequate estimation of these morphological properties depends highly on the acquisition of a precise vascular segmentation. Computational techniques can help to alleviate the extremely tedious and time-consuming task of manually delineating these structures. However, brain vessel segmentation remains an open problem, specially in challenging scenarios such as bAVMs \cite{babin2013brain, forkert2013computer}. 

As recently stated by Moccia \textit{et al}. \cite{moccia2018blood}, the effectiveness of these methods is determined not only by the algorithm itself but also by the imaging modality, the presence/absence of noise or artifacts, and the anatomical region of interest. Moreover, assessment of the segmentation quality varies according to the task at hand. 
For example, in applications such as simulating the deployment of intrasaccular devices for treating aneurysms \cite{dazeo2021intra, munoz2022simulation}, it is only required to have an accurate segmentation within the aneurysm region. Standard intensity-based segmentation methods---such as thresholding or region growing---achieve good results in this setting because relevant foreground voxels share uniformity.
bAVM studies, on the other hand, require a robust coverage of low-intensity distal vessels and of the noisy entanglement of the nidus and draining veins. 
While non-deep learning-based alternatives have been proposed specifically for 3DRA images with bAVMs, these had either scarce\cite{chenoune2019three} or no\cite{clarenccon2015elaboration,li2014segmentation} quantitative validation on real images, or rely heavily on user input\cite{clarenccon2015elaboration}, thus hampering their reproducibility.

Deep learning approaches have been used on a wide variety of biomedical imaging problems\cite{baldi2018deep, haque2020deep}, including the segmentation of the brain vasculature. However, they are either applied on healthy vasculatures\cite{phellan2017vascular, chen20173d} or in images of subjects suffering from other conditions such as acute stroke and steno-occlusive disease\cite{hilbert2020brave}.  
To the best of our knowledge, no methods have been specifically introduced or designed for bAVM segmentation. We believe this could be due to the difficulties in producing an appropriate training set for this task, as irregular blood flow in bAVMs compromises the consistency of the contrast distribution \cite{borden20063d}. Thus, without enough morphological references to rely on, manual annotation becomes tedious, extremely time-consuming and prone to errors and high inter- and intra-observer variability. On the other hand, this also hampers the performance of standard methods to produce first trustworthy initial guesses to work on, requiring an intense posterior polishing labour. Furthermore, bAVMs are a rare condition, which reduces the number of available images to annotate. As a result, collecting a large enough comprehensive data set with densely annotated scans for training deep models becomes prohibitive.



In this paper we propose to overcome this limitation by introducing the first deep learning approach for segmenting challenging bAVMs on 3DRA images. To this end, we densely annotated all brain vessel structures, including feeding arteries, draining veins and the AVM nidus and fistulas, on a private dataset of 5 volumetric 3DRA scans of patients with bAVMs. The resulting dataset allowed us to perform a first comparative study of segmentation performance in the context of bAVM analysis. By building on top of the self-adapting UNet model introduced by Isensee \textit{et al.}\cite{isensee2021nnu}, we defined a 3D patch-based architecture trained with a combination of a standard segmentation losses and a soft centerline Dice\cite{shit2021cldice} based objective. Despite being trained with a reduced amount of data, our final model shows to be accurate enough to display the regions of interest of the bAVM, consistently finding distal arteries and venous structures that sometimes were not even present in the actual ground truth. Hence, this early model could be used to produce first initial guesses of the vasculature in a more trustworthy way, easing the task of generating manual ground truth labels. Furthermore, it opens the way towards applications such as achieving a better delineation of the nidus drainage, assisting transvenous interventional planning and achieving a larger labelled dataset to train more robust solutions. This line of work holds clinical relevance as it aims towards acquiring a quantitative assessment specific to each patient, which would ultimately allow for a better treatment outcome.


\section{Methods and Materials}

\subsection{Dataset and image acquisition}
\label{sec:dataset}

Our dataset comprises 5 contrast-enhanced 3DRA scans of subjects with bAVM (3 assigned female at birth, with ages ranging from 23 to 55 years), collected at the Changhai Hospital (Naval Medical University, Shanghai, P. R. China) before interventional treatment. Data collection was approved by the clinical review board of this medical institution. 
The images were acquired with a Siemens AXIOM-Artis LEO22541 station with an isometric voxel size of 0.36mm--0.46mm, resulting in high resolution volumes of 397 to 511 slices of $512 \times 512$ pixels each.

\subsection{Segmentation model}

   \begin{figure} [t!]
   \begin{center}
   \begin{tabular}{c} 
   \includegraphics[width=0.97\textwidth]{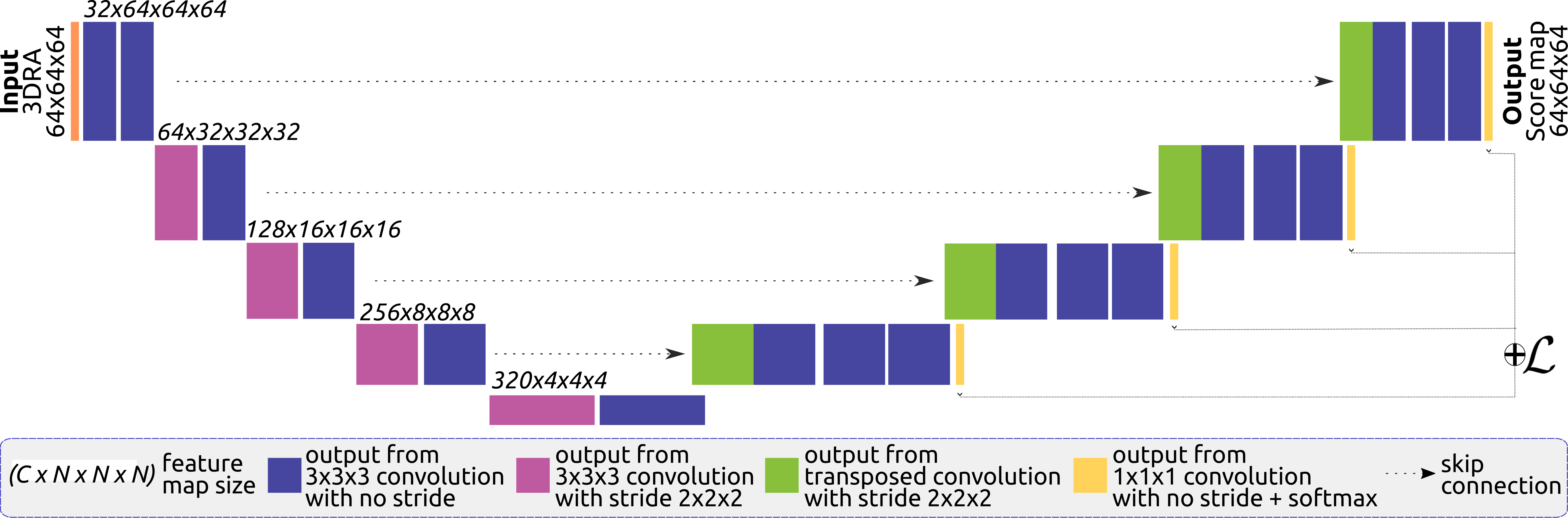}
   \end{tabular}
   \end{center}
   \caption[UNet] 
   { \label{fig:unet} 
   Architecture of the proposed 3DUNet for brain vessel segmentation in bAVM cases. Inputs are patches of size $64 \times 64 \times 64$ pixels, extracted from the normalised 3DRA volume. Downsampling is performed with strided convolutions and upsampling with transposed convolutions. The last four layers of the decoder produce auxiliary segmentation outputs that are used for deep supervision.}
   \end{figure}

\subsubsection{Network architecture}

All five volumes were cropped to the region of interest, removing sections without vasculature. Images were upsampled to have a uniform spacing of 0.227mm across all cases. Volume voxels were also normalised according to the mean and standard deviation of the intensities of each image.

Our network is based on the nnUNet\cite{isensee2021nnu} framework, which allows to find optimal configurations of UNets\cite{ronneberger2015u,cciccek20163d} for biomedical image segmentation by self-adapting them to heterogeneous datasets. We adapted the available source code to train the UNet models with self-configured parameters, such as number of layers and convolutions, as well as patch and batch size. 

Due to the limited amount of samples, we trained five models to segment 3DRA images of bAVMs through a leave-one-out cross-validation \cite{goodfellow2016deep} configuration. Given that our images are volumetric, using the whole angiography as input to the network is computationally prohibitive, as the amount of RAM required to process the produced feature maps exponentially surpassed our hardware disponibility. Additionally, patch-based training is a useful strategy to maximise training data. Hence, to train a 3DUnet architecture as depicted in Figure \ref{fig:unet}, we used patches of $64 \times 64 \times 64$ pixels. 

The encoder in our 3DUNet performs four downsampling operations with strided convolutions, starting with 32 kernels and duplicating them at each layer. The decoder uses transposed convolutions to upsample the input feature maps. Each convolutional layer is followed by instance normalisation and leaky ReLU as a non-linearity. Additional segmentation outputs are obtained from the last four layers for deep supervision\cite{lee2015deeply}.

\subsubsection{Loss function}
Two different objectives were used for training, seeking the most adequate approach to segment the vessel structures in images with bAVMs.
Firstly, we followed a combined cross-entropy and Dice based loss approach \cite{taghanaki2019combo}, namely:
\begin{equation*}
 \mathcal{L}_\text{combo}  = (1-\alpha)\mathcal{L}_\text{CE} + \alpha\mathcal{L}_\text{Dice},
\end{equation*}
\noindent where $\mathcal{L}_\text{CE}$ represents the cross-entropy loss, $\mathcal{L}_\text{Dice}$ the soft Dice loss, and $\alpha \in [0,1]$ the weighted contribution of the Dice term to $\mathcal{L}_\text{combo}$. The Dice loss is widely used in the context of image segmentation, particularly in presence of a class-imbalanced problem. As Dice is undefined when the target of interest is not present in the image, the cross-entropy loss is added as an extra term to penalize errors in these cases and to provide gradient stability. Furthermore, notice that, as the optimal value of Dice coefficient is $1$ (given with absolute prediction-ground truth overlap), its value is internally multiplied by $-1$ to optimise the Dice loss function by its minimum value.

Additionally, we defined a second loss function which includes a centerline-focused metric: 
\begin{equation*}
    \mathcal{L}_\text{combo+clDice} =(1-\alpha)\mathcal{L}_\text{CE} + \alpha((1-\beta)\mathcal{L}_\text{Dice} + \beta\mathcal{L}_\text{clDice}),
\end{equation*}
where $\mathcal{L}_\text{clDice}$ is a soft centerline Dice loss function\cite{shit2021cldice} that favours overlap based on a centerline approximation and vascular connection, instead of sparse voxel overlap. The soft centerline Dice coefficient is internally multiplied by $-1$ to obtain its loss value. $\beta \in [0,1]$ is the weighted contribution of $\mathcal{L}_\text{clDice}$ to the overlap term. We implemented this function to work on 3D images, obtaining soft-skeletons based on pooling operations.

\subsubsection{Experimental setup}
   \begin{figure} [t!]
   \begin{center}
   \begin{tabular}{c} 
   \includegraphics[width=0.87\textwidth]{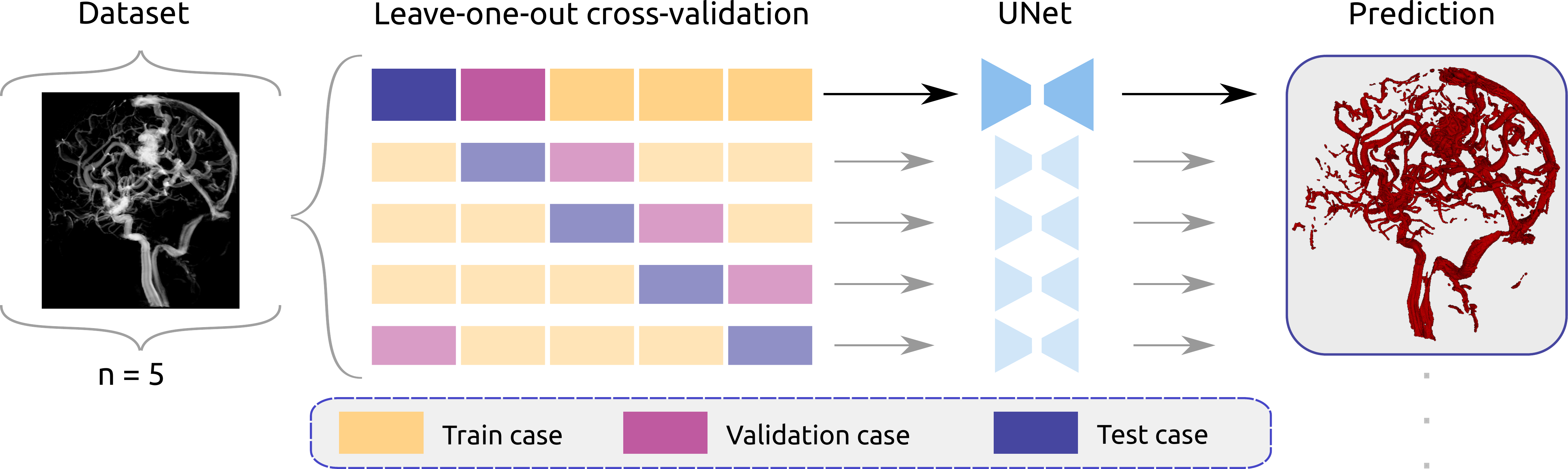}
   \end{tabular}
   \end{center}
   \caption[whatsthis] 
   { \label{fig:workflow} 
Schematic overview of the leave-one-out cross-validation training. Each case in turn is reserved to act as test set. The network is trained then over the remaining four: three cases serving as training set and the last one as validation. The result is set of five 3DUNets trained over our five annotated cases.}
   \end{figure} 
   
In a leave-one-out cross-validation fashion as shown in Figure \ref{fig:workflow}, each of the five models was trained using a fixed batch size of 2 $64 \times 64 \times 64$ pixels patches randomly extracted from a set of 4 cases (3 volumes serving as train set and 1 as validation). Initial learning rate of 0.01, stochastic gradient descent with Nesterov momentum ($\mu = 0.99$) for 250 epochs and loss weights $\alpha=0.5$ and $\beta=0.5$.
Test case prediction was done patch-wise using a sliding window with 50\% overlap, using Gaussian-window weights\cite{isensee2021nnu} to merge all score maps into a single volume and reduce stitching artifacts. Final binarisation of the score map was done with a fixed threshold of 0.5.
The models were trained on a station with 32GB RAM, Intel Core i5-8400 processor and using a GPU NVIDIA GeForce RTX 3060 12GB.

\subsection{Data annotation}

Each study was retrospectively processed to produce reliable ground truth annotations of the brain vasculature for both training and evaluating the proposed model. First initial guesses of the structures of interest were obtained by semi-automatically segmenting the major arteries using empirically defined regions (similar to octree subdivisions) and threshold values (adjusted individually to each region). These annotations were heavily corrected by hand by an engineer with cerebral angiographies analysis expertise, using the segmentation editor provided by 3D Slicer\cite{fedorov20123d}. The correction protocol consisted of adding missing branching vessels, separating merged vessels, segmenting the AVM nidus and covering the venous drainage. The entire manual processing took approximately 10-12 hours for each case. Exemplary ground truth labels are shown in Figures \ref{fig:qualiwhole} and \ref{fig:qualidetails}.

\section{Results}
\label{sec:results}

We evaluated the models both quantitatively and qualitatively, to account for the different aspects which could be of interest considering the further study of the obtained segmentations.

\subsection{Quantitative results}

\begin{table}[t!]
\caption{Quantitative results to a partial ablation study for each of the 5 cases over our two models: $\text{3DUNet}_\text{combo}$ trained with $\mathcal{L}_\text{combo}$ and $\text{3DUNet}_\text{combo+clDice}$ trained with $\mathcal{L}_\text{combo+clDice}$ , evaluated in terms of Dice, Recall (Re) and Precision (Pr) for all vessels and their centerline approximations.}
\label{tab:metricstwo}
\begin{center}       
\begin{tabular}{cl|c|c|c|c|c|c}
\hline
Case & \multicolumn{1}{c}{Model} & \multicolumn{3}{c}{Vessel} & \multicolumn{3}{c}{Centerline} \\
\hline
  \multicolumn{1}{c}{} & \multicolumn{1}{c}{} & \multicolumn{1}{c}{Dice} & \multicolumn{1}{c}{Re} & \multicolumn{1}{c}{Pr} & \multicolumn{1}{c}{Dice} & \multicolumn{1}{c}{Re} & \multicolumn{1}{c}{Pr} \\
\hline
\hline
\multirow{2}{*}{Case 1} & $\text{3DUNet}_\text{combo}$ & \textbf{0.83} & 0.82 & \textbf{0.87} & \textbf{0.87} & 0.90 & \textbf{0.85}  \\ 
                        & $\text{3DUNet}_\text{combo+clDice}$ & \textbf{0.83} & \textbf{0.84} & 0.82 & 0.86 & \textbf{0.91} & 0.82  \\
\hline
\multirow{2}{*}{Case 2} & $\text{3DUNet}_\text{combo}$ & \textbf{0.84} & 0.82 & \textbf{0.86} & \textbf{0.89} & \textbf{0.91} & 0.88  \\ 
                         & $\text{3DUNet}_\text{combo+clDice}$ & \textbf{0.84} & \textbf{0.83} & 0.85 & \textbf{0.89} & 0.88 & \textbf{0.90}     \\
\hline
\multirow{2}{*}{Case 3} & $\text{3DUNet}_\text{combo}$ & \textbf{0.82} & 0.86 & \textbf{0.78} & 0.82 & 0.94 & 0.73  \\ 
                         & $\text{3DUNet}_\text{combo+clDice}$ & 0.80 & \textbf{0.88} & 0.74 & \textbf{0.85} & \textbf{0.95} & \textbf{0.76}  \\
\hline
\multirow{2}{*}{Case 4} & $\text{3DUNet}_\text{combo}$ & \textbf{0.74} & \textbf{0.98} & \textbf{0.59} & 0.76 & \textbf{0.99} & 0.62  \\ 
                         & $\text{3DUNet}_\text{combo+clDice}$ & 0.72 & \textbf{0.98} & 0.57 & \textbf{0.78} & \textbf{0.99} & \textbf{0.64} \\
\hline
\multirow{2}{*}{Case 5} & $\text{3DUNet}_\text{combo}$ & \textbf{0.70} & 0.68 & \textbf{0.71} & \textbf{0.72} & \textbf{0.77} & \textbf{0.67}  \\ 
                         & $\text{3DUNet}_\text{combo+clDice}$ & \textbf{0.70 }& \textbf{0.70} & 0.70 & 0.71 & \textbf{0.77} & 0.66  \\
\hline
\end{tabular}
\end{center}
\end{table}

\begin{table}[t!]
\caption{Quantitative results obtained over all our cases using thresholding, region growing and our two 3DUNet models trained with $\mathcal{L}_\text{combo}$ ($\text{3DUNet}_\text{combo}$) and $\mathcal{L}_\text{combo+clDice}$ ($\text{3DUNet}_\text{combo+clDice}$), evaluated in terms of mean $\pm$ standard deviation to Dice, Recall (Re) and Precision (Pr) for all vessels and for their centerline approximations.} 
\label{tab:metrics}
\begin{center}       
\begin{tabular}{l|c|c|c|c|c|c}
\hline
\multicolumn{1}{c}{Method} & \multicolumn{3}{c}{Vessel} & \multicolumn{3}{c}{Centerline} \\
\hline
  \multicolumn{1}{c}{} & \multicolumn{1}{c}{Dice} & \multicolumn{1}{c}{Re} & \multicolumn{1}{c}{Pr} & \multicolumn{1}{c}{Dice} & \multicolumn{1}{c}{Re} & \multicolumn{1}{c}{Pr} \\
\hline
\hline
\rule[-1ex]{0pt}{3.5ex}  Threshold & $0.66\pm0.17$ & $0.67\pm0.28$ & $0.78\pm0.14$ & $0.62\pm0.19$ & $0.66\pm0.33$ & $0.74\pm0.18$ \\
\hline
\rule[-1ex]{0pt}{3.5ex}  Region Growing & $0.63\pm0.22$ & $0.52\pm0.24$ & $\textbf{0.93}\pm0.03$ & $0.62\pm0.29$ & $0.52\pm0.29$ & $\textbf{0.95}\pm0.03$     \\
\hline
\rule[-1ex]{0pt}{3.5ex}  $\text{3DUNet}_\text{combo}$ & $\textbf{0.78}\pm0.05$ & $0.83\pm0.09$ & $0.76\pm0.09$ & $0.80\pm0.06$ & $\textbf{0.90}\pm0.07$ & $0.75\pm0.10$ \\
\hline
\rule[-1ex]{0pt}{3.5ex}  $\text{3DUNet}_\text{combo+clDice}$ & $0.77\pm0.05$ & $\textbf{0.85}\pm0.09$ & $0.74\pm0.09$ & $\textbf{0.81}\pm0.06$ &$\textbf{0.90}\pm0.07$ & $0.76\pm0.09$  \\
\hline
\end{tabular}
\end{center}
\end{table}

Quantitative results for each case are shown in Table \ref{tab:metricstwo}, obtained using the proposed model trained with $\mathcal{L}_\text{combo+clDice}$ ($\text{3DUNet}_\text{combo+clDice}$) and its counterpart trained using only $\mathcal{L}_\text{combo}$ ($\text{3DUNet}_\text{combo}$). For this partial ablation study, both are evaluated using the overall structures of interest (vessels) and at centerline level in terms of Dice, Recall (Re) and Precision (Pr). Centerline Dice values were computed based on scikit-image\cite{van2014scikit} skeletonisations. 

Average metrics over all our cases are presented in Table \ref{tab:metrics}, including those obtained with two alternative and standard segmentation methods, namely thresholding and region growing. To favour the reproducibility of the experiments, we used the ITK 4.13.0\cite{itk} implementations. Threshold values were set by firstly computing the Otsu threshold and correcting it empirically to reduce the amount of included noise. Region growing was computed by manually selecting initial seeds inside the internal carotid artery, iterating twice with a statistical membership criterion across an 8-connected neighbourhood with a standard deviation multiplier of 2.    




\subsection{Qualitative results}

\begin{figure} [t!]
   \begin{center}
   \begin{tabular}{c} 
   \includegraphics[width=0.95\textwidth]{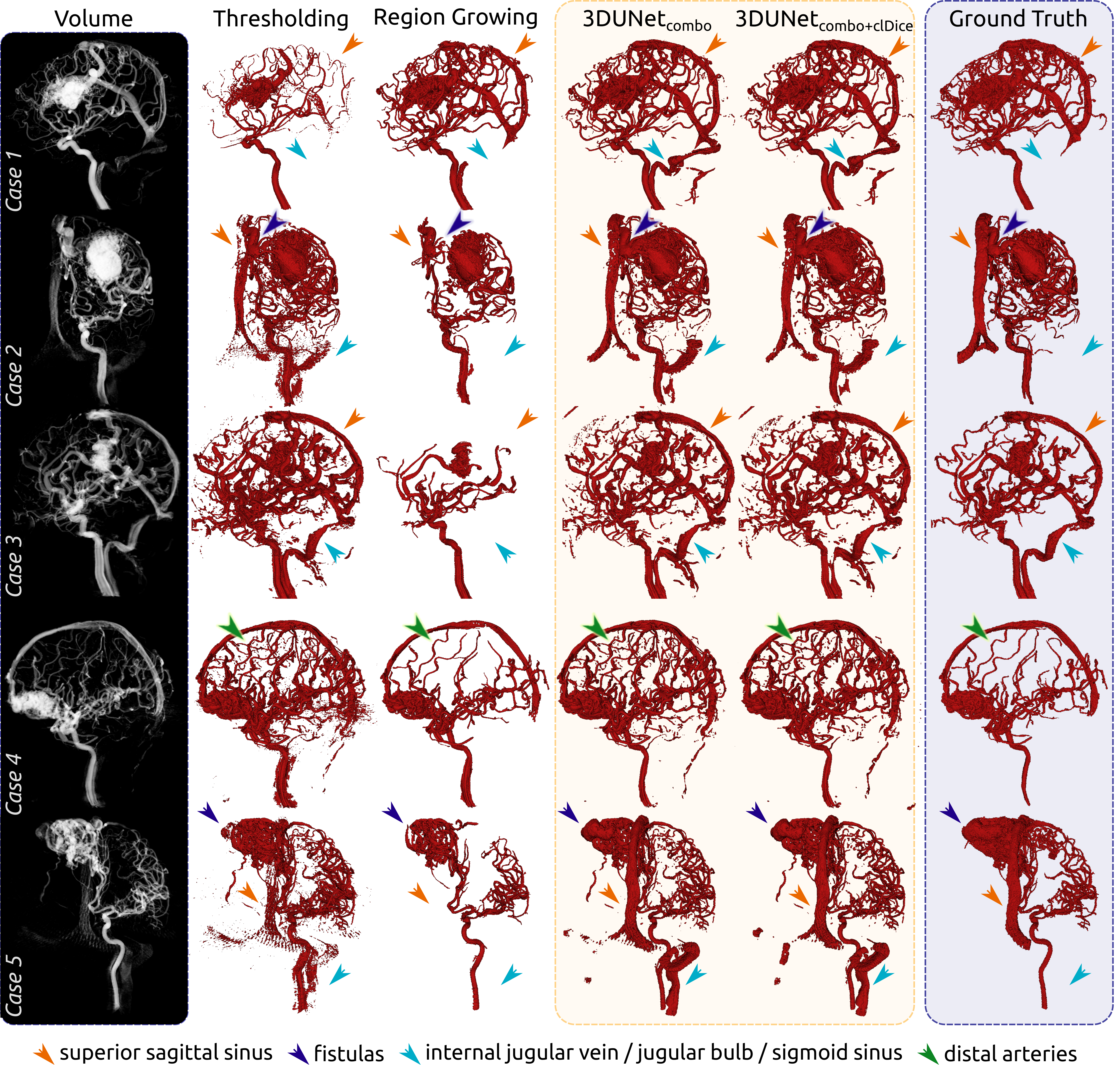}
   \end{tabular}
   \end{center}
   \caption[whatsthis] 
   { \label{fig:qualiwhole} 
Qualitative results for each of the five cases. First column shows a volume rendering of the corresponding case, followed by the segmentations achieved by thresholding, region growing, $\text{3DUNet}_\text{combo}$ and $\text{3DUNet}_\text{combo+clDice}$. Lastly, the manual annotations acting as ground truth. Volume projections correspond to sagittal or coronal orientations, according to the AVM location and clearest segmentation overview.}
   \end{figure}

\begin{figure} [t!]
   \begin{center}
   \begin{tabular}{c} 
   \includegraphics[width=0.97\textwidth]{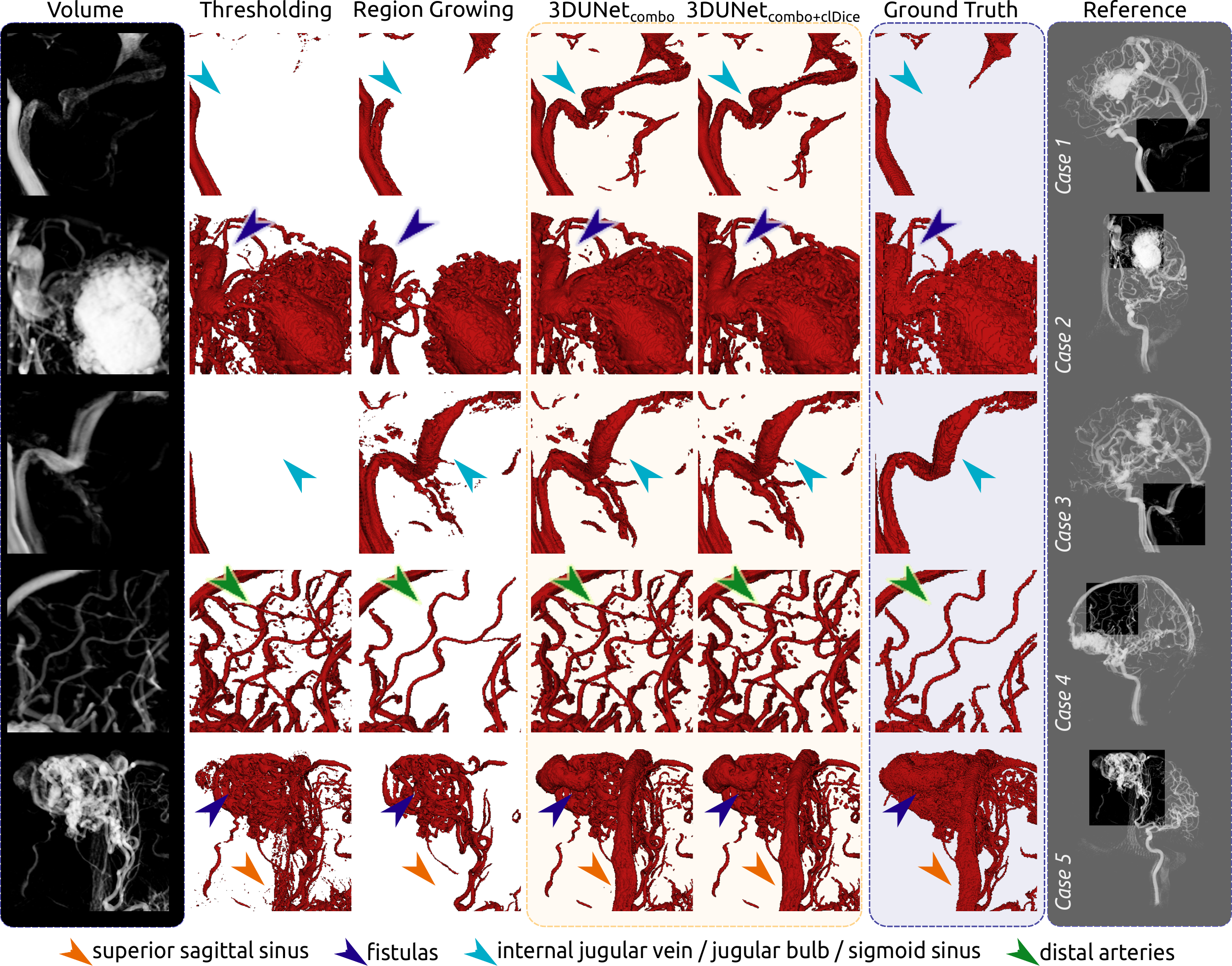}
   \end{tabular}
   \end{center}
   \caption[whatsthis] 
   { \label{fig:qualidetails} 
Enlarged view of the qualitative results presented in Figure~\ref{fig:qualiwhole}. Top to bottom: vein of case 1, fistula and venous drainage of case 2, vein structure in case 3, distal arteries to case 4, dense nidus and draining veins to case 5.}
   \end{figure}

In a general overview, our 3DUNet models reached a higher discovery rate of the overall labelled vessel structures. In Figure \ref{fig:qualiwhole} we present projections of the segmentations achieved by the methods on each of the five cases, and some relevant regions are shown enlarged in Figure \ref{fig:qualidetails}. 
For most cases, thresholding and region growing missed draining vessels and had an insufficient coverage of venous structures present in ground truth. For example: ground truth for cases 1, 2 and 5 did not include some vein structures (like the jugular vein or sigmoid sinus) that were segmented by the models; cases 1, 2, 3 and 5 show that our models found the superior sagittal sinus while the standard segmentation methods missed it, just like the fistulas in cases 2 and 5; many distal arteries were also segmented by the models but were not present in the ground truth, as can be observed in case 4.


\section{Discussion}
\label{sec:discussion}
We approached the segmentation of 3DRA images of bAVMs with deep learning methods after densely annotating 5 volumes of a private dataset. Our aim was to acquire sufficiently good binarisations that allow a posterior topological and morphological analysis of bAVMs. The assessment was based on a quantitative comparison of two overlap metrics: vessel Dice coefficient and centerline Dice coefficient. These were selected to account for the overall relevant voxels and also preservation of voxel connectivity.

We performed a partial ablation study to understand the effect of incorporating the additional centerline-focused loss during training. No training was carried on with only cross-entropy loss, given the class-imbalance inherent to brain vessel images. 
Method comparison was narrowed down to threshold and region growing, which are standard in vessel segmentation tasks and their code implementations are readily available. 
This study design decision was deemed appropriate to minimise reproducibility difficulties and potential ambiguity. 
There are limitations to this approach, since our network works patch-wise while threshold and region growing were used volume-wise. 
However, using these standard methods on a patch level would imply threshold level and seed computation for each patch, which would then introduce additional complexity to the comparison.

Regarding the quantitative analysis, we considered vessel Dice and centerline Dice coefficients with their respective Precision and Recall values. Our UNet models consistently achieve a higher Dice coefficient value and Recall value with respect to the standard segmentation methods, indicating a better coverage of the foreground voxels. Even though Precision values are higher with region growing, the resulting segmentation is insufficient in the overall recovery of foreground voxels.

Methodologically, we observed that incorporating the additional centerline-focused loss during training does not offer an evident improvement in the quantitative results with respect to the combined cross-entropy and Dice losses approach (Table \ref{tab:metrics} and Table \ref{tab:metricstwo}).
Nevertheless, qualitative analysis hints towards clearer vessel delineations and enhanced connectivity when employing soft centerline Dice loss calculation (as in case 4 in Figure \ref{fig:qualidetails}).
Furthermore, considering centerline Dice as a form of quantitative report provides additional reassurance regarding preservation of vessel-like qualities \cite{shit2021cldice}.

False positives yielded by our models correspond mostly to veins and distal arteries that were not delineated in ground truth (see Figure \ref{fig:qualiwhole}). We consider this aspect presents two potential and interconnected advantages. 
Firstly, acquiring ground truth labels for bAVMs angiographies has proven to be a challenge, due to the lack of enough available images, high inter-rater variability \cite{moccia2018blood} and unpredictable topomorphology \cite{dumont2017brain, rutledge2017brain}. 
Manually generating these labels is an intensely time-consuming task, and even carefully polished segmentations are prone to present missing vessels. 
We found that models trained over densely annotated volumes were able to yield adequate binarisations with high structure coverage.
It would then be possible to leverage these segmentations to ease the manual annotation task, by using the predicted geometries as first initial guesses of the structures of interest.
Additionally, appropriately dealing with the segmentation of venous structures is a difficult task due to shape and intensity inconsistencies. 
High flow rates in bAVM cases can also trigger vascular remodeling, which further compromises venous delineation by rendering statistical atlas unreliable. 
Nonetheless, it is becoming highly relevant for clinical practice as interventional treatments start to consider and involve transvenous embolisation more often \cite{chen2018transvenous,chen2020brain,waldeck2021first}. 
As preliminary approaches, skeleton-based vein extractions have been proposed \cite{babin2018skeletonization} given a sufficiently good vascular segmentation. Hence, developing a model able to handle these difficulties at segmentation time could be a valuable tool for embolisation planning in clinical practice. 

The presented work constitutes a further step towards achieving a robust segmentation method despite the mentioned difficulties. This is only the first part of the still challenging task of comprehensive technology-assisted bAVM study, but crucial to its achievement. 
Our results are encouraging regarding satisfactory coverage of all structures of interest, and might eventually help alleviate the burdensome manual labelling task or even serve as a starting point to reach more reliable segmentations. 
Ultimately, this would contribute to a better understanding and achieving a more accurate morphological and topological characterisation of these malformations. 
In this respect, further tests are currently underway to assess the impact on model robustness of the inclusion of diverse data augmentation techniques and synthetic vessel datasets.

\appendix    

\acknowledgments 

CG is funded by a CONICET PhD Scholarship. This work was partially funded by PICT startup 2021-00023 - FONCYT - ANPCYT (Argentina), PIP 2021-2023 11220200102472CO from CONICET (Argentina) and a NVIDIA Hardware Grant. It was also partially supported by the computing facilities of Extremadura Research Centre for Advanced Technologies (CETA-CIEMAT), funded by the European Regional Development Fund (ERDF). CETA-CIEMAT belongs to CIEMAT and the Government of Spain.

\bibliography{report} 

\begin{thebibliography}{10}

\bibitem{dumont2017brain}
Dumont, A.~S. et~al.,  [{\em Brain Arteriovenous Malformations and
  Arteriovenous Fistulas}{\nolinebreak\hspace{0.1em}]}, Thieme (2017).

\bibitem{rutledge2017brain}
Rutledge, W.~C. and Lawton, M.~T.,  [{\em Brain AVM: Current Treatments and
  Challenges}{\nolinebreak\hspace{0.1em}]}, ch.~4,  69--81 (2017).

\bibitem{rutledge2014hemorrhage}
Rutledge, W.~C. et~al., ``Hemorrhage rates and risk factors in the natural
  history course of brain arteriovenous malformations,'' {\em Translational
  stroke research}~{\bf 5}(5),  538--542 (2014).

\bibitem{spetzler1986proposed}
Spetzler, R.~F. and Martin, N.~A., ``A proposed grading system for
  arteriovenous malformations,'' {\em Journal of neurosurgery}~{\bf 65}(4),
  476--483 (1986).

\bibitem{lawton2010supplementary}
Lawton, M.~T. et~al., ``A supplementary grading scale for selecting patients
  with brain arteriovenous malformations for surgery,'' {\em Neurosurgery}~{\bf
  66}(4),  702--713 (2010).

\bibitem{tranvinh2017contemporary}
Tranvinh, E. et~al., ``Contemporary imaging of cerebral arteriovenous
  malformations,'' {\em AJR}~{\bf 208}(6),  1320--1330 (2017).

\bibitem{babin2013brain}
Babin, D. et~al., ``Brain blood vessel segmentation using line-shaped
  profiles,'' {\em Physics in Medicine \& Biology}~{\bf 58}(22),  8041 (2013).

\bibitem{forkert2013computer}
Forkert, N.~D. et~al., ``Computer-aided nidus segmentation and angiographic
  characterization of arteriovenous malformations,'' {\em IJCARS}~{\bf 8}(5),
  775--786 (2013).

\bibitem{moccia2018blood}
Moccia, S. et~al., ``Blood vessel segmentation algorithms—review of methods,
  datasets and evaluation metrics,'' {\em Computer methods and programs in
  biomedicine}~{\bf 158},  71--91 (2018).

\bibitem{dazeo2021intra}
Dazeo, N. et~al., ``Intra-saccular device modeling for treatment planning of
  intracranial aneurysms: from morphology to hemodynamics,'' {\em International
  Journal of Computer Assisted Radiology and Surgery}~{\bf 16}(10),  1663--1673
  (2021).

\bibitem{munoz2022simulation}
Mu{\~n}oz, R. et~al., ``Simulation of intra-saccular devices for pre-operative
  device size selection: Method and validation for sizing and porosity
  simulation,'' {\em Computers in Biology and Medicine} ,  105744 (2022).

\bibitem{chenoune2019three}
Chenoune, Y. et~al., ``Three-dimensional segmentation and symbolic
  representation of cerebral vessels on 3dra images of arteriovenous
  malformations,'' {\em Comput. Biol. Med}~{\bf 115},  103489 (2019).

\bibitem{clarenccon2015elaboration}
Claren{\c{c}}on, F. et~al., ``Elaboration of a semi-automated algorithm for
  brain arteriovenous malformation segmentation: initial results,'' {\em
  European radiology}~{\bf 25}(2),  436--443 (2015).

\bibitem{li2014segmentation}
Li, F. et~al., ``Segmentation and reconstruction of cerebral vessels from 3d
  rotational angiography for avm embolization planning,'' in [{\em 2014 36th
  Annual International Conference of the IEEE Engineering in Medicine and
  Biology Society}{\nolinebreak\hspace{0.1em}]},   5522--5525, IEEE (2014).

\bibitem{baldi2018deep}
Baldi, P., ``Deep learning in biomedical data science,'' {\em Annual review of
  biomedical data science}~{\bf 1},  181--205 (2018).

\bibitem{haque2020deep}
Haque, I. R.~I. and Neubert, J., ``Deep learning approaches to biomedical image
  segmentation,'' {\em Informatics in Medicine Unlocked}~{\bf 18},  100297
  (2020).

\bibitem{phellan2017vascular}
Phellan, R. et~al., ``Vascular segmentation in tof mra images of the brain
  using a deep convolutional neural network,'' in [{\em Intravascular Imaging
  and Computer Assisted Stenting, and Large-Scale Annotation of Biomedical Data
  and Expert Label Synthesis}{\nolinebreak\hspace{0.1em}]},   39--46, Springer
  (2017).

\bibitem{chen20173d}
Chen, L. et~al., ``3d intracranial artery segmentation using a convolutional
  autoencoder,'' in [{\em 2017 IEEE International Conference on Bioinformatics
  and Biomedicine (BIBM)}{\nolinebreak\hspace{0.1em}]},   714--717, IEEE
  (2017).

\bibitem{hilbert2020brave}
Hilbert, A. et~al., ``Brave-net: fully automated arterial brain vessel
  segmentation in patients with cerebrovascular disease,'' {\em Frontiers in
  artificial intelligence} ,  78 (2020).

\bibitem{borden20063d}
Borden, N.~M.,  [{\em 3D angiographic atlas of neurovascular anatomy and
  pathology}{\nolinebreak\hspace{0.1em}]}, Cambridge University Press (2006).

\bibitem{isensee2021nnu}
Isensee, F. et~al., ``nnu-net: a self-configuring method for deep
  learning-based biomedical image segmentation,'' {\em Nature methods}~{\bf
  18}(2),  203--211 (2021).

\bibitem{shit2021cldice}
Shit, S. et~al., ``cldice-a novel topology-preserving loss function for tubular
  structure segmentation,'' in [{\em Proceedings of the IEEE/CVF Conference on
  Computer Vision and Pattern Recognition}{\nolinebreak\hspace{0.1em}]},
  16560--16569 (2021).

\bibitem{ronneberger2015u}
Ronneberger, O., Fischer, P., and Brox, T., ``U-net: Convolutional networks for
  biomedical image segmentation,'' in [{\em International Conference on Medical
  image computing and computer-assisted
  intervention}{\nolinebreak\hspace{0.1em}]},   234--241, Springer (2015).

\bibitem{cciccek20163d}
{\c{C}}i{\c{c}}ek, {\"O}. et~al., ``3d u-net: learning dense volumetric
  segmentation from sparse annotation,'' in [{\em International conference on
  medical image computing and computer-assisted
  intervention}{\nolinebreak\hspace{0.1em}]},   424--432, Springer (2016).

\bibitem{goodfellow2016deep}
Goodfellow, I., Bengio, Y., and Courville, A.,  [{\em Deep
  learning}{\nolinebreak\hspace{0.1em}]}, MIT press (2016).

\bibitem{lee2015deeply}
Lee, C.-Y. et~al., ``Deeply-supervised nets,'' in [{\em Artificial intelligence
  and statistics}{\nolinebreak\hspace{0.1em}]},   562--570, PMLR (2015).

\bibitem{taghanaki2019combo}
Taghanaki, S.~A. et~al., ``Combo loss: Handling input and output imbalance in
  multi-organ segmentation,'' {\em Computerized Medical Imaging and
  Graphics}~{\bf 75},  24--33 (2019).

\bibitem{fedorov20123d}
Fedorov, A. et~al., ``3d slicer as an image computing platform for the
  quantitative imaging network,'' {\em Magnetic resonance imaging}~{\bf 30}(9),
   1323--1341 (2012).

\bibitem{van2014scikit}
Van~der Walt, S. et~al., ``scikit-image: image processing in python,'' {\em
  PeerJ}~{\bf 2},  e453 (2014).

\bibitem{itk}
McCormick, M.~M. et~al., ``{ITK}: enabling reproducible research and open
  science,'' {\em Frontiers in neuroinformatics}~{\bf 8},  13 (2014).

\bibitem{chen2018transvenous}
Chen, C.-J. et~al., ``Transvenous embolization of brain arteriovenous
  malformations: a review of techniques, indications, and outcomes,'' {\em
  Neurosurgical focus}~{\bf 45}(1),  E13 (2018).

\bibitem{chen2020brain}
Chen, C.-J. et~al., ``Brain arteriovenous malformations: a review of natural
  history, pathobiology, and interventions,'' {\em Neurology}~{\bf 95}(20),
  917--927 (2020).

\bibitem{waldeck2021first}
Waldeck, S. et~al., ``First experience in the control of the venous side of the
  brain avm,'' {\em Journal of Clinical Medicine}~{\bf 10}(24),  5771 (2021).

\bibitem{babin2018skeletonization}
Babin, D. et~al., ``Skeletonization method for vessel delineation of
  arteriovenous malformation,'' {\em Computers in Biology and Medicine}~{\bf
  93},  93--105 (2018).

\end{thebibliography}
\bibliographystyle{spiebib} 

\end{document}